\documentclass[floats,floatfix,showpacs,amssymb,prd,twocolumn,superscriptaddress,nofootinbib]{revtex4-1}

\usepackage{graphicx,epsf, epsfig, amssymb}
\usepackage{bm}
\usepackage{longtable}
\usepackage{color}
\usepackage[breaklinks]{hyperref}
\usepackage[caption=false]{subfig}
\usepackage{amsfonts,amsmath,amssymb,mathrsfs}

\def\be{\begin{equation}}
\def\ee{\end{equation}}
\def\beq{\begin{eqnarray}}
\def\eeq{\end{eqnarray}}

\newcommand{\bea}{\begin{eqnarray}}
\newcommand{\eea}{\end{eqnarray}}
\newcommand{\ben}{\begin{enumerate}}
\newcommand{\een}{\end{enumerate}}
\newcommand{\bi}{\begin{itemize}}
\newcommand{\ei}{\end{itemize}}

\newcommand{\nn}{\nonumber}

\def\nn{\nonumber}
\def\be{\begin{equation}}
\def\ee{\end{equation}}
\def\beq{\begin{eqnarray}}
\def\eeq{\end{eqnarray}}

\def\th{\vartheta}

\begin{document}

\title{\large Gravito-Electromagnetic Perturbations
  of Kerr-Newman Black Holes:\\ Stability and Isospectrality in the 
Slow-Rotation Limit}

\author{Paolo Pani}\email{paolo.pani@ist.utl.pt}
\affiliation{CENTRA, Departamento de F\'{\i}sica, Instituto Superior
  T\'ecnico, Universidade T\'ecnica de Lisboa - UTL, Avenida~Rovisco Pais
  1, 1049 Lisboa, Portugal.}
\affiliation{Institute for Theory \& Computation, Harvard-Smithsonian
  CfA, 60 Garden Street, Cambridge, MA, USA}

\author{Emanuele Berti}
\affiliation{Department of Physics and Astronomy, The University of
  Mississippi, University, MS 38677, USA.}
\affiliation{California Institute of Technology, Pasadena, CA 91109, USA}

\author{Leonardo Gualtieri}
\affiliation{Dipartimento di Fisica, Universit\`a di Roma ``La
  Sapienza'' \& Sezione INFN Roma1, P.A. Moro 5, 00185, Roma, Italy.}

\date{\today} 

\begin{abstract}
  The most general stationary black-hole solution of Einstein-Maxwell
  theory in vacuum is the Kerr-Newman metric, specified by three
  parameters: mass $M$, spin $J$ and charge $Q$. Within classical
  general relativity, one of the most important and challenging open
  problems in black-hole perturbation theory is the study of
  gravitational and electromagnetic fields in the Kerr-Newman
  geometry, because of the indissoluble coupling of the perturbation
  functions. Here we circumvent this long-standing problem by working
  in the slow-rotation limit. We compute the quasinormal modes up to
  linear order in $J$ for any value of $Q$ and provide the first,
  fully-consistent stability analysis of the Kerr-Newman metric. For
  scalar perturbations the quasinormal modes can be computed exactly,
  and we demonstrate that the method is accurate within $3\%$ for
  spins $J/J_{\rm max}\lesssim 0.5$, where $J_{\rm max}$ is the
  maximum allowed spin for any value of $Q$. Quite remarkably, we find
  numerical evidence that the axial and polar sectors of the
  gravito-electromagnetic perturbations are {\em isospectral} to
  linear order in the spin. The extension of our results to
  nonasymptotically flat space-times could be useful in the context
  of gauge/gravity dualities and string theory.
\end{abstract}

\pacs{04.70.Bw, 04.25.Nx, 04.30.Db}

\maketitle
\date{today}

\noindent{\bf{\em Introduction.}}
In Einstein-Maxwell theory, black holes (BHs) that are stationary,
asymptotically flat end-states of gravitational collapse must be
axisymmetric~\cite{Hawking:1971vc}. Classic uniqueness theorems
reviewed in~\cite{Chrusciel:2012jk} show that regular, stationary
electrovacuum BH space-times in four dimensions are described by the
Kerr-Newman (KN) metric~\cite{Newman:1965my}, characterized by mass
$M$, angular momentum $J$ and electromagnetic charge $Q$. When $Q=0$
the KN solution reduces to the Kerr metric, and for $J=0$ it reduces
to the Reissner-Nordstr\"om (RN) metric. When both $Q$ and $J$ are
nonvanishing the space-time is endowed with an induced magnetic field,
and its magnetic dipole moment corresponds to the same gyromagnetic
ratio $g=2$ as the electron~\cite{Carter:1968rr}.  This observation
led to some speculation that the KN metric could be used as a
classical model for elementary particles (see
e.g.~\cite{Pekeris:1989mu}).

Charge is unlikely to play a significant role in
astrophysics~\cite{Gibbons:1975kk,Blandford:1977ds}, but the KN metric
is still a precious theoretical laboratory to investigate Einstein-Maxwell theory in curved space-time. 
For this reason the linearized dynamics of
test fields on a KN background have been intensively studied in the
past. The scalar~\cite{Dadhich:2001sz}, neutrino~\cite{Unruh:1973},
massive spin-$1/2$~\cite{Chandrasekhar:1976ap,Page:1976jj} and
Rarita-Schwinger~\cite{TorresdelCastillo:1990aw} equations in the KN
metric can all be solved by separation of variables.
The scattering of charged scalars and fermions in near-extremal KN
space-times recently acquired special interest in the context of the
KN/Conformal Field Theory (CFT)
conjecture~\cite{Hartman:2008pb,Hartman:2009nz}.

The KN space-time is one of the simplest prototypes of the interplay
between matter and curvature summarized by Wheeler's famous statement
that ``matter tells space-time how to curve, and space-time tells
matter how to move.''  Despite their importance, theoretical
investigations of the interplay between gravitational and
electromagnetic perturbations in the KN metric are still in their
infancy. The reason is a major technical stumbling block: most methods
to compute quasinormal modes (QNMs, see
\cite{Kokkotas:1999bd,Nollert:1999ji,Berti:2009kk,Konoplya:2011qq} for
reviews), greybody factors and scattering amplitudes rely on
separability, and despite several
attempts~\cite{Dudley:1977zz,Dudley:1978vd,Bellezza:1984}, at present
no one has been able to separate the angular and radial dependence of
the gravito-electromagnetic eigenfunctions\footnote{The last chapter
  of Chandrasekhar's monumental 1983 monograph \cite{Chandra} is
  dedicated to an incomplete treatment of this problem. Quoting
  from~\cite{Chandra}: ``It does not appear that the methods developed
  [...]  for the treatment of the gravitational perturbations of the
  Kerr black-hole can be extended in any natural way to the treatment
  of the coupled electromagnetic and gravitational perturbations of
  the KN black-hole. The origins of this apparently essential
  difference in the perturbed Kerr and KN space-time may lie
  deep in the indissoluble coupling of the spin-1 and spin-2 fields in
  the perturbed KN space-time -- a coupling which it was
  possible to break only for very special reasons in the perturbed
  RN space-time.''}.

This work is the first consistent analysis of gravito-electromagnetic
perturbations of KN BHs. We circumvent the decades-old coupling
problem using a recent framework to study generic perturbations of
spinning BHs in the slow-rotation
limit~\cite{Pani:2012vp,Pani:2012bp}, which is based on a similar
approach used in the past to study slowly rotating compact stars
\cite{Kojima:1992ie,1993ApJ...414..247K,ChandraFerrari91,Ferrari:2007rc}.
We summarize here some of our most interesting findings: (i) We
present the first self-consistent calculation of scalar,
electromagnetic and gravitational QNMs of the KN metric. (ii) Since
none of these modes is unstable, our calculation provides solid
evidence for the stability of the (nonextremal) KN metric. (iii) In
the scalar case we can compare our results to an exact calculation,
that does not rely on the slow-rotation limit. We find that the
perturbative analysis is valid when $J/J_{\rm max}\ll 1$ (where
$J_{\rm max}$ is the maximum allowed spin for any given $Q$) and that
scalar QNM frequencies are accurate within $3\%$ for spins $J/J_{\rm
  max}\lesssim 0.5$, which suggests a similar level of accuracy for
the gravito-electromagnetic modes. (iv) Last but not least, we find
the remarkable result that axial and polar QNMs (corresponding to
perturbations that have odd or even parity, respectively) are {\em
  isospectral} to linear order in the spin.

\noindent{\bf{\em Formalism.}}
The KN metric is the most general stationary electrovacuum solution of
Einstein-Maxwell theory. Its full form in Boyer-Lindquist coordinates
can be found, e.g., in~\cite{Chandra}. Here and in the following we
linearize all quantities in the spin parameter $\tilde a\equiv a/M\equiv J/M^2$ (in
geometrical units $G=c=1$), neglecting terms of order ${\cal O}(\tilde
a^2)$. To this order, the KN metric reads
\begin{align}
ds^2_0=-Fdt^2 
+F^{-1}dr^2-2\varpi\sin^2\th d\varphi dt+r^2d^2\Omega\,,
\label{metric0}
\end{align}
where $F(r)=(r-r_-)(r-r_+)/r^2$, $r_\pm=M\pm\sqrt{M^2-Q^2}$ 
are the horizons of a RN BH, the gyromagnetic term is
\begin{eqnarray}
\varpi(r)&=&{2\tilde{a} M^2}/{r}-{\tilde{a}Q^2M}/{r^2},
\end{eqnarray}
and the background electromagnetic potential is given by
\begin{equation}
A_\mu=\left(\frac{Q}{r},0,0,-\frac{\tilde{a}Q 
M}{r}\sin^2\vartheta\right)\,.
\end{equation}
Note that the presence of \emph{both} rotation and charge ($\tilde a 
Q\neq0$) induces a magnetic field in the $(\vartheta,\varphi)$ 
directions.

We derive the equations describing gravito-electromagnetic
oscillations in the slow-rotation approximation by linearizing the
Einstein-Maxwell equations with respect to both the oscillation
amplitude and the BH spin parameter $\tilde{a}$, and by expanding the
perturbations in a complete basis of tensor spherical harmonics.
As a consequence of using this basis in a nonspherical background, the
linearized equations display mixing between perturbations with
different harmonic indices and opposite
parity~\cite{ChandraFerrari91,Kojima:1992ie,Pani:2012vp,Pani:2012bp}. However,
the latter do not contribute to the QNM spectrum to first order in
$\tilde a$ \cite{1993ApJ...414..247K,1993PThPh..90..977K,Pani:2012bp}.

Our main analytical result consists of two sets of coupled,
second-order equations (one for the axial and one for the polar
sector, respectively) which fully describe gravito-electromagnetic
oscillations of a KN BH to first order in the spin. In the
frequency-domain, and assuming a time dependence $e^{-i\omega t}$,
they read (schematically)
\begin{eqnarray}
\hat {\cal D} Z_i^{\pm}&&\equiv V_0^{(i,\pm)}Z_i^{\pm}+m \tilde a\left[V_1^{(i,\pm)}Z_i^{\pm}+V_2^{(i,\pm)}{Z_i^{\pm}}'\right]\nn\\
&&+m \tilde a Q^2\left[W_1^{(i,\pm)} Z_j^{\pm}+W_2^{(i,\pm)}{Z_j^{\pm}}'\right],\label{eF}
\end{eqnarray}
where $i,j=1,2$, $i\neq j$ (there is no sum over the indices $i,j$),
we have defined the differential operator $\hat{\cal D}=\partial^2_{r_*}+\omega^2-F{\ell(\ell+1)}/{r^2}$, and $r_*$ is the
standard tortoise coordinate, such that $\partial_{r_*}r=F$. The functions
$Z_i^{-}$ and $Z_i^{+}$ are linear combinations of axial and polar
variables, respectively, and they are also combinations of
gravitational and electromagnetic perturbations.

The explicit form of the axial and polar potentials $V^{(i,\pm)}$ and
$W^{(i,\pm)}$ is quite formidable. It will be presented in the
accompanying paper~\cite{PaniPRD} and in a publicly available
{\scshape Mathematica} notebook~\cite{webpage}.  What matters is that
Eqs.~\eqref{eF} display the same symmetries as the master equations
for a RN BH~\cite{Chandra}, and indeed they exactly reduce to the
latter in the nonrotating case.
In addition, Eqs.~\eqref{eF} contain two first-order corrections in
$\tilde{a}$. The first term 
is responsible for a Zeeman-like splitting of the eigenfrequencies,
which breaks the degeneracy in the azimuthal index $m$. The second
line in Eq.~\eqref{eF} is more interesting: this term couples the
function $Z_1^+$ with the function $Z_2^+$, and the function $Z_1^-$
with the function $Z_2^-$. 

Once physically-motivated boundary conditions are imposed,
Eq.~\eqref{eF} defines an eigenvalue problem for the complex frequency
$\omega=\omega_R+i\omega_I$.  The boundary conditions for QNMs read
simply
\begin{equation}
 Z_j^\pm(r)\sim \left\{\begin{array}{l}
                        e^{{i}\omega r_*},\qquad \hspace{1.35cm} r\to\infty\\
                        e^{-{i} (\omega-m\Omega_H) r_*},\qquad r\to r_+
                       \end{array}\right.
\label{asymp}
\end{equation}
The near-horizon solution displays the typical frame-dragging effect
occurring near spinning BHs, where
\begin{equation}
\Omega_H\sim\frac{\tilde{a}}{M (1+{\tilde a}_{\rm 
max})^2}+{\cal O}(\tilde{a}^3)\label{OmegaH}
\end{equation}
is the angular velocity at the horizon of locally nonrotating
observers, and $\tilde a_{\rm max}\equiv J_{\rm
  max}/M^2=\sqrt{1-(Q/M)^2}$ is the maximum spin parameter of a KN BH.

\begin{figure}[thb]
\begin{center}
\begin{tabular}{c}
\epsfig{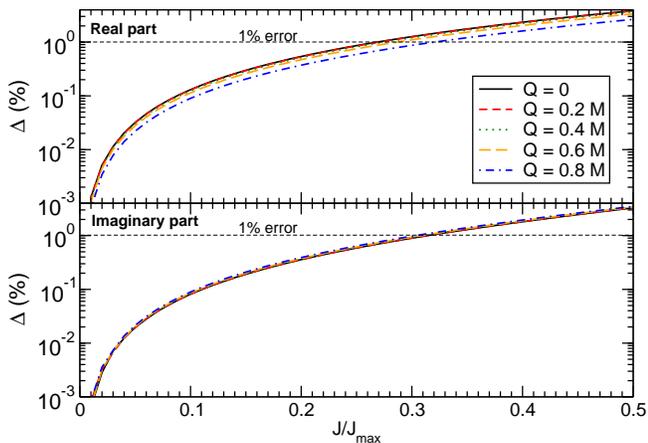}
\end{tabular}
\caption{Top panel: percentage error $\Delta\equiv 10^2|1-\omega_{\rm
    slow}/\omega_{\rm exact}|$ induced by the slow-rotation
  approximation in the real part of the fundamental $\ell=m=1$ scalar
  mode. Bottom panel: percentage error for the imaginary part of the
  same mode. The errors are only mildly sensitive to $Q$ if plotted as
  functions of $J/J_{\rm max}$, where $J_{\rm max}$ is defined below 
Eq.~\eqref{OmegaH}. Similar results hold for other scalar 
modes~\cite{PaniPRD}.
\label{fig:scalar}}
\end{center}
\end{figure}
%

\noindent{\bf{\em Numerical Results.}}
The numerical solution of the axial and polar perturbation
equations~\eqref{eF} is challenging, because their explicit form is
very complicated~\cite{PaniPRD,webpage}. Robust numerical methods to
solve the coupled eigenvalue problem given by Eqs.~\eqref{eF} with the
boundary conditions~\eqref{asymp} are
reviewed in~\cite{PaniNRHEP2}. We have integrated the coupled
system~\eqref{eF} and computed the corresponding eigenfrequencies
using two independent methods: a highly-efficient matrix-valued
continued fraction technique and direct integration~\cite{PaniPRD,PaniNRHEP2}.
When both methods are applicable they validate each other, in the
sense that the results agree within numerical accuracy.

For any given $Q$, our analysis allows us to extract the first-order
corrections to the complex QNM frequencies:
\begin{eqnarray} 
\omega_{R,I}&=&\omega_{R,I}^{(0)}+\tilde{a}m\,\omega_{R,I}^{(1)}+{
\cal O}(\tilde{a}^2),\label{wRI}
\end{eqnarray}
where $\omega_{R,I}^{(i)}$ are functions of $Q$ and of the multipolar 
index $\ell$, whereas the $m$-dependence has been factored 
out~\cite{Pani:2012bp}.

As a test of the slow-rotation approximation, we have computed the
scalar QNMs of a KN BH to first order in $\tilde{a}$. These modes can
be computed exactly in the Teukolsky
formalism~\cite{Berti:2005eb}, so they give us the precious
opportunity to estimate the errors introduced by the slow-rotation
approximation. 
For any stationary and axisymmetric space-time, the
scalar modes at first order in the angular momentum are governed by a
master equation~\cite{Pani:2012bp} whose corresponding eigenvalue 
problem can be solved with standard 
continued-fraction techniques~\cite{Leaver:1990zz,Berti:2004md}. 

In Fig.~\ref{fig:scalar} we show the relative error of the
slow-rotation approximation with respect to the ``exact'' result,
computed by solving the scalar equation in a KN background via
continued fractions~\cite{Berti:2005eb}. In particular, the top
(bottom) panels show the percentage deviation $\Delta$ for the real
(imaginary) part of the fundamental $\ell=m=1$ scalar mode at fixed
values of $Q$. The slow-rotation
approximation is accurate within one percent as long as $J/J_{\rm
  max}\lesssim 0.3$, and it is still accurate within $3\%$ for
$J/J_{\rm max}\lesssim 0.5$. Similar results also hold for other 
values of $\ell$ and $m$, and for the first few 
overtones~\cite{PaniPRD}.
Note the near-universal behavior of the
percentage errors as functions of $J/J_{\rm max}=\tilde a/{\tilde
  a}_{\rm max}$ for all values of $Q$. Indeed the parameter ${\tilde
  a}_{\rm max}$, which appears explicitly in the QNM boundary
conditions~\eqref{asymp}, plays a fundamental role in our
perturbative scheme and the slow-rotation approximation is accurate
only far from extremality, i.e. when $\tilde a\ll{\tilde a}_{\rm
  max}=\sqrt{1-{Q^2}/{M^2}}<1$.

%
\begin{figure}[thb]
\begin{center}
\begin{tabular}{c}
\epsfig{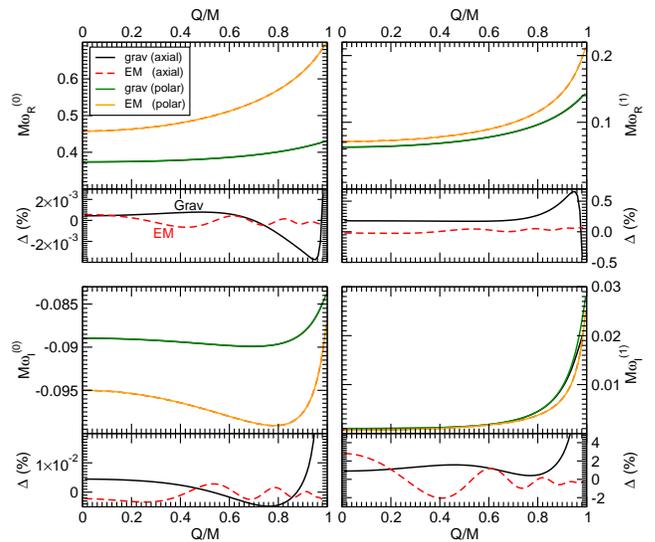}
\end{tabular}
\caption{Zeroth-order (left panels) and first-order (right panels)
  terms of the slow-rotation expansion of the KN QNM frequencies
  [cf.~Eq.~\eqref{wRI}]. All quantities are plotted as a function of
  $Q/M$, and they refer to the fundamental mode ($n=0$) with
  $\ell=2$.
  The lower part of each panel shows the percentage difference between
  axial and polar quantities: our results are consistent with
  isospectrality to ${\cal O}(0.1\%)$ for the real part and to ${\cal
    O}(1\%)$ for the imaginary part of these modes.
\label{fig:modes}}
\end{center}
\end{figure}

Figure~\ref{fig:modes} shows our main numerical results for the
fundamental gravito-electromagnetic perturbations with $\ell=2$, the
most relevant for gravitational-wave emission (see
e.g.~\cite{Berti:2005ys}). In each panel we show four curves,
corresponding to the axial and polar ``gravitational'' and
``electromagnetic'' modes (as defined in the decoupled limit,
$Q\to0$).  The zeroth-order terms shown in the left panels are simply
RN QNMs; they agree with continued-fraction
solutions~\cite{Leaver:1990zz} of the equations first derived by
Zerilli~\cite{Zerilli:1974ai}.

We carried out a more extensive QNM calculation working in the axial
case, where our results can be verified using two independent
methods. By virtue of the isospectrality between axial and polar modes
visible in Fig.~\ref{fig:modes} and discussed below, these results
cover the whole QNM spectrum of slowly rotating KN BHs.
We found that the zeroth- and first-order quantities shown in
Fig.~\ref{fig:modes} (plus analogous calculations for $\ell>2$ and 
the first overtones~\cite{PaniPRD}) are well fitted by functions of 
the form
\begin{equation}
 \omega^{(0,1)}_{R,I}= 
\sum_{k=0}^4 f_k y^k\,.
\label{fit}
 \end{equation}
Here we have defined a parameter $y=1-{\tilde a}_{\rm max}$, which is
in one-to-one correspondence with $Q/M$, but is better suited for
fitting. The coefficients $f_i$ 
for the fundamental $\ell=2$
gravito-electromagnetic modes are listed in Table~\ref{tab:fit}.
%
\begin{table}[htb]
 \begin{tabular}{cc|ccccc}
  &	($\ell$,$n$,$s$)	& $f_0$& $f_1$& $f_2$& $f_3$& $f_4$ \\
\hline
$\omega_R^{(0)}$ & (2,0,1) &  0.4576 &  0.2659 &  0.0118 &  0.1228 & -0.1382 \\
$\omega_R^{(1)}$ & (2,0,1) &  0.0712 &  0.0769 &  0.0596 &  0.0727 & -0.0216 \\
$\omega_I^{(0)}$ & (2,0,1) & -0.0950 & -0.0184 &  0.0137 &  0.0132 &  0.0107 \\
$\omega_I^{(1)}$ & (2,0,1) &  0.0007 &  0.0043 &  0.0060 & -0.0089 &  0.0366 \\
\hline
$\omega_R^{(0)}$ & (2,0,2) &  0.3737 &  0.0525 &  0.0607 & -0.0463 & -0.0070 \\
$\omega_R^{(1)}$ & (2,0,2) &  0.0628 &  0.0676 &  0.0209 &  0.0823 & -0.0810 \\
$\omega_I^{(0)}$ & (2,0,2) & -0.0890 & -0.0055 &  0.0024 &  0.0214 & -0.0084 \\
$\omega_I^{(1)}$ & (2,0,2) &  0.0010 &  0.0014 &  0.0091 &  0.0174 &  0.0145 \\
\hline
\hline
\end{tabular}
\caption{Coefficients of the fit~\eqref{fit} for the real and
  imaginary part of the fundamental ($n=0$) gravito-electromagnetic
  modes with $\ell=2$. 
  We denote by $s=1$ and $s=2$ the modes that in the decoupled
  limit, $Q\to0$, are electromagnetic and gravitational,
  respectively.  The fits~\eqref{fit} reproduce the data to within
  $1\%$ for $\omega_I^{(1)}$, and to within $0.1\%$ for the other
  quantities, for any $Q\lesssim0.95 M$. Similar fits have comparable
  accuracy also for $\ell>2$ and for the first overtone
 ~\cite{PaniPRD}.
\label{tab:fit}}
\end{table} 

\noindent{\bf{\em Stability.}}
None of our numerical searches (for $0<Q<M$, $J\ll J_{\rm max}$ and
$\ell=2,3,4$) returned exponentially growing modes. This confirms
early arguments by Mashhoon in favor of the stability of the KN
metric~\cite{Mashhoon:1985}. Mashhoon's results apply only to the
eikonal limit ($\ell\gg 1$) and they rely on a somewhat heuristic
geodesic analogy, rather than on a self-consistent treatment of the
perturbation equations. In this sense, our QNM calculations provide
the first, fully consistent numerical evidence for the stability of
the KN space-time.

\noindent{\bf{\em Isospectrality.}}
Gravito-electromagnetic perturbations of Schwarzschild and RN BHs in
general relativity have a noteworthy property, first proved by
Chandrasekhar~\cite{Chandra}: even though the polar and axial sectors
of the perturbations are described by completely different potentials,
their QNM spectra are identical~\cite{Berti:2009kk}. Mathematically,
this happens because the polar and axial potentials can be written in
terms of a single ``superpotential'' (cf. Sections~26 and 43 of
\cite{Chandra}). This property can be interpreted as being due to
``supersymmetry,'' in the sense of nonrelativistic quantum mechanics
\cite{Witten:1981nf,Cooper:1994eh,2001JMP....42.4802L}.

Isospectrality is easily broken: e.g., it does not hold if the
cosmological constant is nonzero
\cite{Mellor:1989ac,Cardoso:2001bb,Berti:2003ud}, if the underlying
theory is not general relativity~\cite{Cardoso:2009pk,Molina:2010fb},
or in higher dimensions (cf.~Appendix A
of~\cite{Berti:2009kk}). The left panels of Fig.~\ref{fig:modes}
(which refer to the RN limit) show that polar and axial modes are
isospectral within our numerical accuracy. Given the complex form of
the polar equations, this is a nontrivial check of our numerical
techniques.

A priori, there is no reason why such a remarkable and fragile
symmetry should hold true also for rotating (KN) BHs.  A
tantalizing result of our numerical study is strong evidence that the
axial and polar sectors of KN gravito-electromagnetic perturbations
are indeed isospectral to first order in the BH spin. The left panels
of Fig.~\ref{fig:modes} show that the linear corrections
$\omega_{R,I}^{(1)}$ are identical functions of $Q$ for axial and
polar modes within the numerical errors (which are dominated by
uncertainties in the direct integration used to compute polar modes).
Various arguments can be made to support the claim that the observed
deviations from isospectrality are of a purely numerical nature: 1)
isospectrality is verified to a higher level of accuracy far from
extremality: this is consistent with the fact that QNMs are more 
challenging to compute in the extremal limit; 2) the deviations
from isospectrality shown in Fig.~\ref{fig:modes} are
roughly constant or decreasing functions of $Q$ (at least for
$Q\lesssim 0.8M$) and they are affected by a small residual error even
when $Q=0$, where isospectrality must hold exactly;
3) the direct integration method 
is more accurate as $\ell$ grows and, correspondingly, 
the deviations between axial and polar modes decrease;
4) finally, we verified that the error can be reduced by
increasing the accuracy of the integrator.

\noindent{\bf{\em Outlook.}}
It is tempting to conjecture that the isospectrality we found at
linear order may in fact hold exactly, at all orders in
rotation. In order to verify this hypothesis it will be crucial to
include effects of second order in the spin -- a formidable
undertaking.  At second order the causal structure of a spinning
metric starts differing from the nonspinning case, and parity-mixing
terms appear in the perturbation equations~\cite{Pani:2012bp}. 
If isospectrality were to
hold true also at second order, there would be no fundamental reason
to believe that it should be broken at higher orders. However, let us
stress that isospectrality is a highly nontrivial property even at
linear order in rotation, in view of the mixing of gravitational and
electromagnetic perturbations.
Hopefully our work will stimulate further studies to verify whether
isospectrality is an exact property of the KN space-time. Besides
brute-force extensions of our work to higher orders in rotation, other
possible means of studying this problem include numerical time
evolutions (cf.~\cite{Dorband:2006gg,Witek:2012tr,Dolan:2012yt}) or
analytical work, perhaps along the lines of
\cite{2001JMP....42.4802L}.

Other interesting extensions of this work concern asymptotically anti-de Sitter (AdS) space-times.  
Our approach can be easily applied to compute the QNMs of
(slowly rotating) KN-AdS BHs. This would be useful in the context of
the AdS/CFT correspondence~\cite{Maldacena:1997re}, which predicts
that these solutions are dual to thermal states of a CFT living in a
rotating Einstein universe~\cite{Caldarelli:1999xj,Hawking:1999dp}.

Such an extension would also be interesting in the context of
supergravity. To clarify this point, let us recall that the QNMs of
asymptotically flat, extremal RN BHs have a remarkable property:
electromagnetic perturbations with angular index $\ell$ are
isospectral with gravitational perturbations with index $\ell+1$
\cite{Onozawa:1995vu}. This is related to the fact that, when
embedded in $N=2$, four-dimensional supergravity, these solutions
preserve part of the supersymmetry. Using this
property, it is possible to prove that the one-loop corrections to the
BH entropy cancel~\cite{Kallosh:1997ug}.  In the case of rotating, asymptotically flat BHs
this reasoning does not apply, since these solutions are not
supersymmetric. However,
certain KN-AdS BHs embedded in $N=2$ four-dimensional supergravity,
preserve half of the supersymmetry~\cite{Kostelecky:1995ei,Caldarelli:1998hg}. 
Since these solutions can be
slowly rotating, the techniques developed in this paper could be
used to extend the arguments of~\cite{Kallosh:1997ug} to
supersymmetric KN-AdS BHs.

%
Last but not least, it would be interesting to extend our calculation
of KN QNMs in the context of the KN/CFT
conjecture~\cite{Hartman:2008pb,Hartman:2009nz}, which predicts that
the QNMs of the near-horizon KN geometry correspond to the poles of
the retarded Green's function of the dual chiral CFT~\cite{Chen:2010i}.

\noindent{\bf{\em Acknowledgments.}}
We thank Vitor Cardoso for useful discussions. This work was supported
by the NRHEP--295189 FP7--PEOPLE--2011--IRSES Grant, and by FCT -
Portugal through PTDC projects FIS/098025/2008, FIS/098032/2008,
CERN/FP/123593/2011. E.B. was supported by NSF CAREER Grant
No.~PHY-1055103. P.P. acknowledges financial support provided by the
European Community through the Intra-European Marie Curie contract
aStronGR-2011-298297.

\bibliography{KN}
\end{document}